\documentclass[twocolumn,aps,prb,amsmath,floatfix]{revtex4}
\usepackage{graphicx}
\begin{document}

\title{Mott transition in multi-orbital systems}
\author{A. Liebsch}
\affiliation{Institut f\"ur Festk\"orperforschung, Forschungszentrum
             J\"ulich, 52425 J\"ulich, Germany}
\date{\today }
\maketitle

{\bf
Metal insulator transitions driven by local Coulomb interactions 
are among the most fascinating phenomena in condensed matter physics
\cite{mott}.
They occur in a large variety of transition metal compounds
\cite{imada}.
Most of these strongly correlated materials consist of valence bands 
derived from electronic $d$ shells where intra- and inter-orbital Coulomb 
interactions are equally important and where the crystal structure splits 
the valence bands into narrow and wide subbands. A fundamental question 
is whether these systems exhibit a common Mott transition, implying 
all subbands to be either metallic or insulating, or successive orbital 
dependent transitions, implying a coexistence region with metallic 
and insulating behavior present in different subbands. Using the 
dynamical mean field theory 
\cite{DMFT} 
we show that inter-orbital Coulomb interactions lead to a single Mott 
transition. Nevertheless, the subbands exhibit more or less strongly 
correlated excitation spectra in the metallic phase and different band 
gaps in the insulating phase.
}

A hallmark of strongly correlated materials is their geometrical,
electronic and magnetic complexity. The valence bands typically
involve oriented electronic orbitals, giving rise to a variety of 
highly anisotropic properties. For instance, the layer perovskite 
Sr$_2$RuO$_4$ consists of a wide, two-dimensional $d_{xy}$ band 
coexisting with narrow, nearly one-dimensional $d_{xz,yz}$ bands 
\cite{oguchi}. 
This system exhibits unconventional p-wave superconductivity  
\cite{maeno},
but iso-electronic replacement of Sr by Ca induces a Mott transition 
to an antiferromagnetic insulator  
\cite{nakatsuji}.   
Similar splittings into different subbands occur in the classic Mott 
insulators VO$_2$\ \cite{goodenough,zylbersztejn}
and V$_2$O$_3$\    \cite{rozenberg,ezhov}
in both the low-temperature insulating (monoclinic) and 
high-temperature metallic (rutile, corundum) phases, 
in layered organic superconductors  
\cite{lefebvre},
fullerenes
\cite{takenobu},
and many other compounds  
\cite{imada}.

The nature of the paramagnetic metal insulator transition in a 
multi-band environment involving narrow and wide subbands is not yet
understood; in particular, it is not clear whether all bands undergo a common transition
at the same critical Coulomb energy 
\cite{EPL} 
or whether different subbands generate transitions at successive critical 
energies  
\cite{anisimov}. 
In the first scenario narrow subbands force wide subbands to become 
more strongly correlated and the more pronounced metallicity of wider 
subbands makes narrow bands less correlated than in the isolated case. 
The critical Coulomb energy $U_c$ of the interacting system lies between 
the critical $U_{ci}$ of the isolated subbands. 
Coexistence of metallic and insulating behavior in different subbands 
does not occur. 
In the second scenario the different relative importance of Coulomb 
correlations in subbands of different widths gives rise to separate 
Mott transitions and to coexistent metallic and insulating behavior 
in different subbands. In this case the one-electron properties 
imposed by the different band widths remain a determining factor for 
the system's excitation spectrum and are not entirely superseded by 
correlations. 
    
While local Coulomb interactions in multi-band systems have been 
investigated previously 
\cite{multi,florens},  
most studies treated the special `isotropic' case of identical orbitals 
where the question of one or several Mott transitions does not arise. 
Here we consider the opposite `non-isotropic' situation where the 
multi-band system consists of  different subbands 
\cite{prl}.  
In the absence of inter-orbital Coulomb 
interactions these bands would exhibit Mott transitions at 
different critical energies $U_{ci}$. As shown below, the striking 
consequence of inter-orbital interactions is to enforce one common
metal insulator transition for all subbands at the same critical $U_c$.
Remarkably, however, in the joint metallic region different subbands 
reveal excitation spectra with varying correlation signatures.          
Similarly, the insulating phase exhibits a `multi-gap' spectrum, 
in close analogy to the $\sigma$- and $\pi$-type superconducting gaps 
observed in the quasi-two-dimensional compound MgB$_2$ 
\cite{MgB2}.

Let us consider the paramagnetic metal insulator transition in a 
two-band Hubbard model with narrow and wide subbands. These might 
represent, for instance, the $t_{2g}$ orbitals of a layer-perovskite 
material or of VO$_2$. For illustrative purposes these bands are assumed 
to be half-filled and to have elliptical densities of states 
$\rho_i(\omega)$ of widths $W_1=2$\,eV and $W_2=4$\,eV. 
Single-band systems of this type have been investigated extensively in 
the past and their metal insulator transitions are rather well 
understood
\cite{single}. 
To account for Coulomb correlations we use the dynamical mean field 
theory (DMFT) in combination with the multi-orbital Quantum Monte Carlo 
(QMC) method
\cite{DMFT}.
The quantum impurity problem is solved at finite temperature $T$ for 
various intra- and inter-orbital Coulomb energies $U$ and $U'=U-2J$, 
where $J$ is the Hund's rule exchange integral.

Two criteria are employed to determine the stability of the Fermi liquid
state as a function of Coulomb energy. First, we calculate the subband 
quasi-particle weights 
\,$Z_i = 1/(1-\partial{\rm Re}\,\Sigma_i(\omega)/\partial\omega
            \vert_{\omega=0})$,
where the derivative of the real part of the self-energy component 
$\Sigma_i(\omega)$ is approximated by \,Im\,$\Sigma_i(i\omega_0)/\omega_0$ 
with $\omega_0=\pi/\beta=\pi k_BT$ the first Matsubara frequency. The second 
criterion is obtained from the imaginary-time Green's functions at 
$\tau=\beta/2$:
\,$G_i(\beta/2) = \int\! d\omega\,F(\omega)\,{\rm Im}\,G_i(\omega)/\pi$,
where \,$F(\omega) = 0.5/{\rm cosh}(\beta\omega/2)$\, is a distribution 
of width \,$w=4\,{\rm ln}(2+\sqrt3)/\beta$\, centered about the Fermi level.
While $Z_i$ specifies the quasi-particle weight of the i$^{th}$ 
subband at $E_F$, $G_i(\beta/2)$ represents the integrated spectral 
weight within a few $k_BT$ of $E_F$, i.e., it includes low-lying 
excitations. 
Since $\Sigma_i(i\omega_n)$ and $G_i(\tau)$ are derived directly 
from the QMC calculation $Z_i$ and $G_i(\beta/2)$ are available
without having to evaluate the real-frequency spectral distributions
\,{\rm Im}\,$G_i(\omega)$. Below we present normalized quantities 
\,$\bar G_i(\beta/2)= G_i(\beta/2)/G_i^{U=0}(\beta/2)$ so that 
$Z_i=\bar G_i(\beta/2)=1$ in the non-interacting limit.  

\begin{figure}[t]
  \begin{center}
  \includegraphics[width=5.5cm,height=8cm,angle=-90]{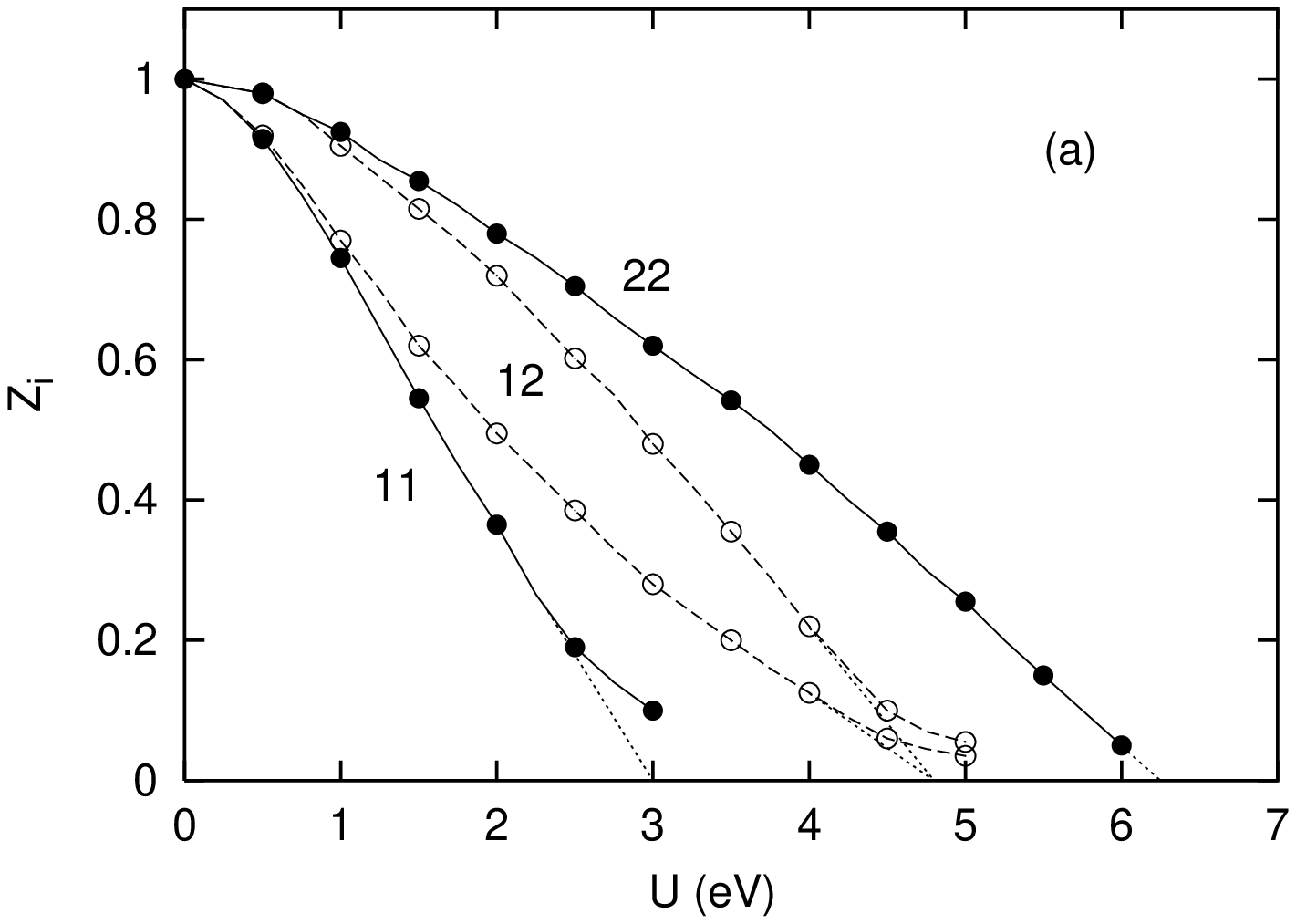}
  \includegraphics[width=5.5cm,height=8cm,angle=-90]{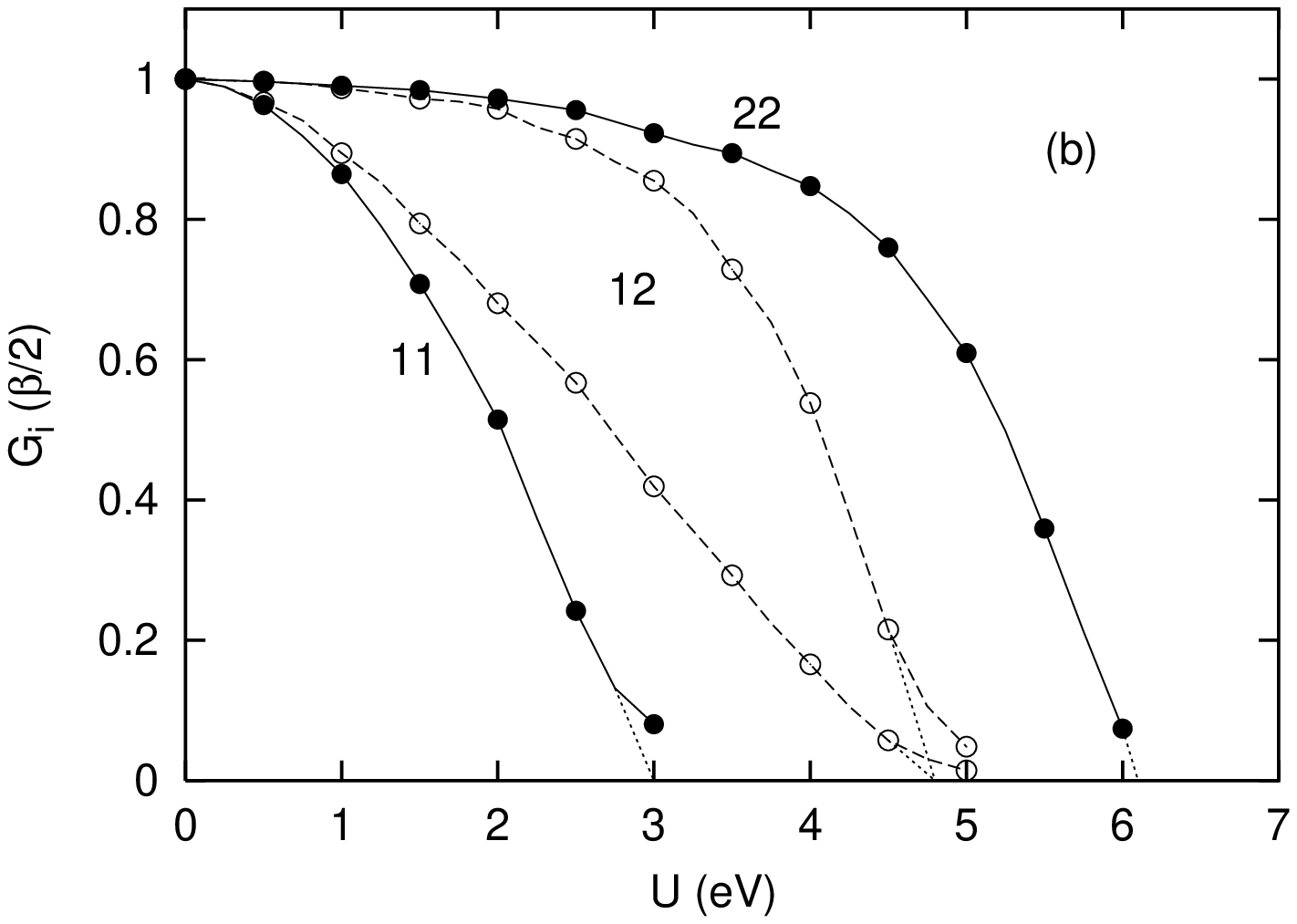}
  \end{center}
\caption{
(a) Quasi-particle weights $Z_i$ and (b) normalized imaginary-time 
Green's functions $\bar G_i(\beta/2)$ of two-band systems as a function 
of $U$ for $T=125$\,meV. Solid symbols: isotropic $W_1,W_1$ and $W_2,W_2$ 
systems; open symbols:  non-isotropic $W_1,W_2$ system. 
Dotted lines: extrapolation of $Z_i$ and $\bar G_i(\beta/2)$
to estimate critical Coulomb energies.
}\end{figure}
    
Figure 1 shows the variation of $Z_i$ and $\bar G_i(\beta/2)$ as a 
function of $U$ for $T=125$\,meV. The exchange integral is  
$J=0.2$\,eV     
\cite{Jbelow1}. 
Evidently the quasi-particle weights near $E_F$ diminish as a result 
of Coulomb correlations. At a given $U$, however, $Z_1$ and $\bar G_1$ 
in the non-isotropic $W_1,W_2$ system 
are larger than in the degenerate $W_1,W_1$ case. Similarly, $Z_2$ and 
$\bar G_2$  in the $W_1,W_2$ system are smaller than in the $W_2,W_2$ 
case. Thus, as a consequence of inter-orbital interactions the narrow 
(wide) subband is less (more) correlated than in the isotropic case.

\begin{figure}[t]
  \begin{center}
  \includegraphics[width=5.5cm,height=8cm,angle=-90]{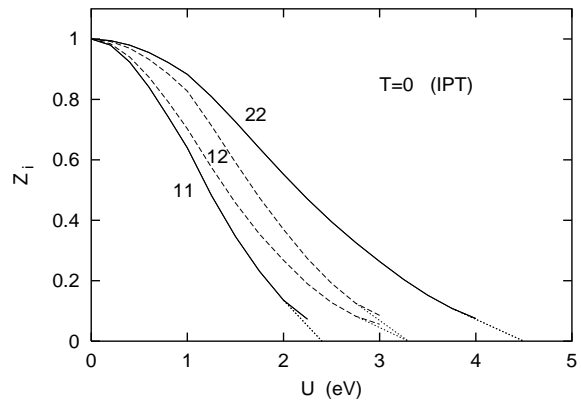}
  \end{center}
\caption{
Quasi-particle weights $Z_i$ of two-band systems as a function of $U$, 
calculated within iterated perturbation theory at $T=0$. 
Solid curves: isotropic $W_1,W_1$ and $W_2,W_2$ systems; 
dashed curves: non-isotropic $W_1,W_2$ system. 
Dotted lines: extrapolation of $Z_i$ to estimate critical Coulomb 
energies.
}\end{figure}

Since $\bar G_i(\beta/2)$ represents the spectral weight within several 
$k_BT$ of $E_F$ its reduction at small $U$ is weaker than that of $Z_i$, 
while in the critical regions its decay is more abrupt. Close to the metal 
insulator transitions $Z_i$ and $\bar G_i$ show a 
rounding off caused by the finite temperature 
and by critical slowing down. Within this slight uncertainty $Z_i$ and 
$\bar G_i$ yield consistent critical Coulomb energies:
$U_{c1}\approx3.0$~eV for the narrow two-band system, 
$U_{c2}\approx6.1$~eV for the wide   two-band system, and 
$U_{c }\approx4.8$~eV for the mixed system 
\cite{factor}.

The key point of these results is the fact that the system involving 
narrow and wide subbands exhibits a single Mott transition at an 
intermediate $U_c$ such that $U_{c1} < U_c < U_{c2}$. At small $U$
correlations reduce the quasi-particle weight in the narrow 
subband much more rapidly than in the wide band. For $U>U_c/2$, however,
$Z_1$ and $Z_2$ begin to converge again and decrease to zero spectral 
weight at the same $U_c$. The same behavior is shown by the integrated 
weights $\bar G_i(\beta/2)$. There is no evidence for a coexistence of 
insulating and metallic behavior in different subbands 
\cite{Uc1Uc2}.  
On the other hand, since $Z_1<Z_2$ and $\bar G_1<\bar G_2$ 
for $U<U_c$, the narrow band is always 
more strongly correlated than the wider counterpart. Analogously,  
the insulating phase exhibits two excitation gaps (see below).

\begin{figure}[t]
  \begin{center}
  \includegraphics[width=5.5cm,height=8cm,angle=-90]{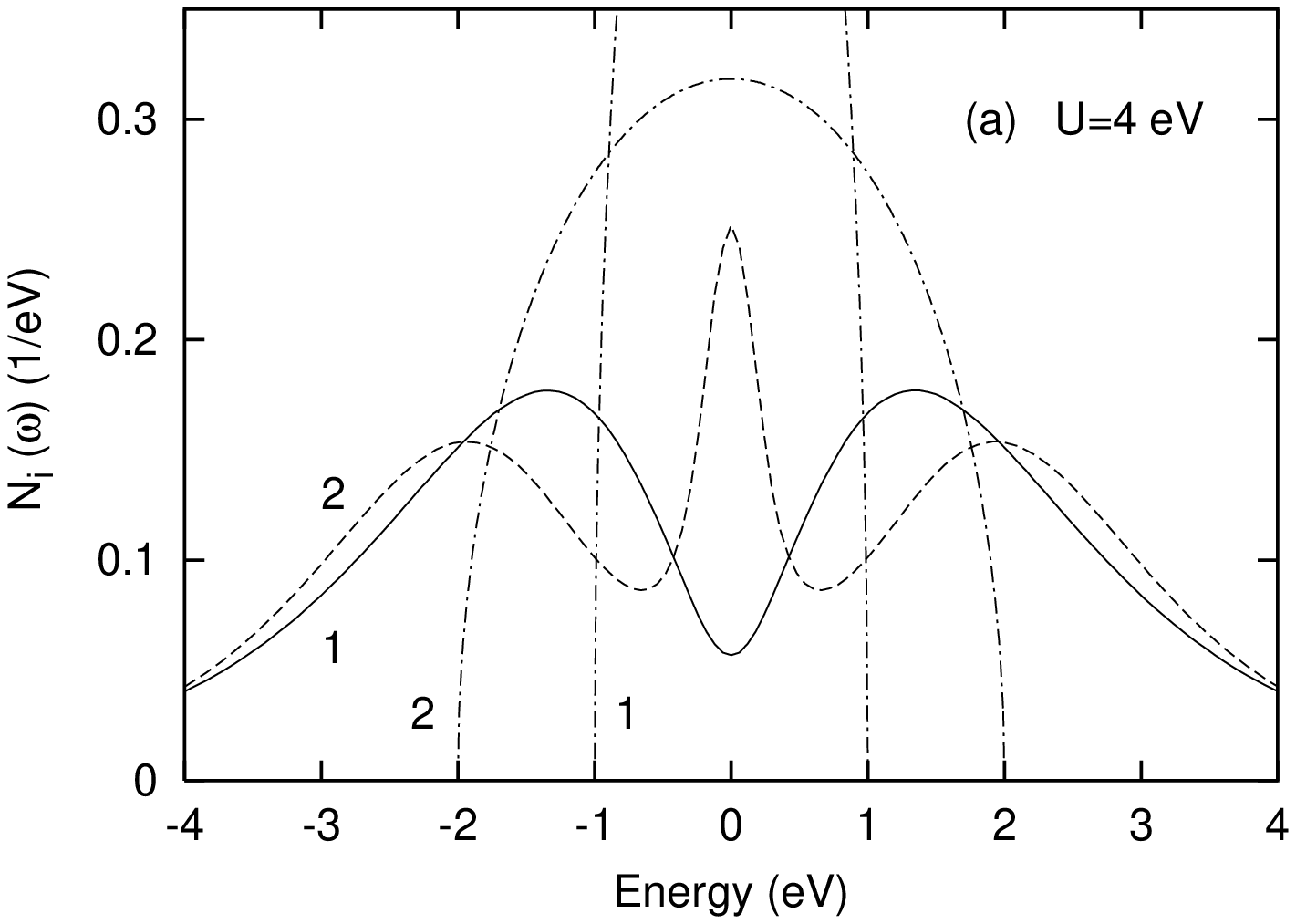}
  \includegraphics[width=5.5cm,height=8cm,angle=-90]{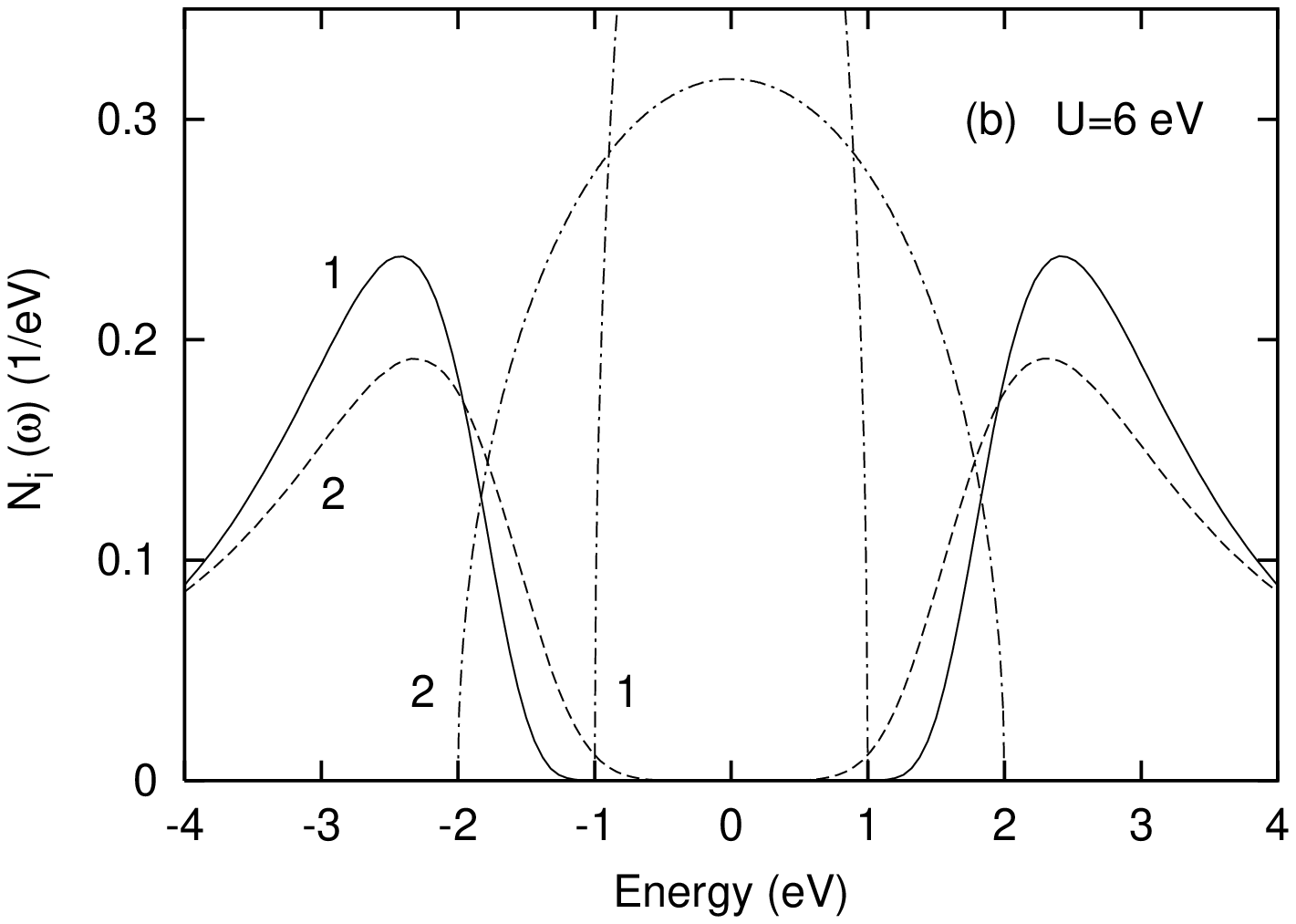}
  \end{center}
\caption{
Quasi-particle spectra of non-isotropic two-band system 
for $T=125$\,meV ($E_F=0$):
(a) metallic phase, (b) insulating phase. 
The solid and dashed curves denote the spectra of the narrow and wide
subbands, respectively. The bare densities of states are 
represented by the dot-dashed curves.
}\end{figure}

Multi-band QMC calculations at lower temperatures ($T=62$\,meV and 31\,meV) 
confirm the picture discussed above. To verify that the same conclusion holds 
also in the low temperature limit we have performed analogous two-band 
calculations at $T=0$ within the self-consistent iterated perturbation theory 
\cite{IPT}
(IPT), where the self-energy is calculated at real frequencies 
within second-order perturbation theory. The results for the quasi-particle 
weights $Z_i$ are shown in Fig.~2. They are consistent with the 
scenario for $T>0$: The isotropic two-band systems 
have widely different critical Coulomb energies:  
$U_{c1}\approx2.3$~eV  and  $U_{c2}\approx4.3$~eV.  
In contrast, the non-isotropic system undergoes a Mott 
transition at $U_{c}\approx3.2$~eV. 
The specific values of these Coulomb energies differ from those 
in Fig.~1 since the IPT amounts to a more approximate treatment of 
correlations, and because of the different temperatures.

Although inter-orbital Coulomb interactions lead to a common metal 
insulator transition in the narrow and wide subbands, their excitation 
spectra differ qualitatively both in the metallic and insulating phases. 
This is illustrated in Fig.~3 where we plot quasi-particle
distributions derived using the Maximum Entropy method
\cite{maxent}.  
For $U<U_c$ both bands are metallic. The narrow band spectrum 
$N_1(\omega)$, however, is more strongly correlated than $N_2(\omega)$ 
for the wider band: The spectral weight at $E_F$ is 
much reduced and the Hubbard peaks are more pronounced.  
Similarly, in the insulating phase the excitation gap of the narrow band 
is larger than that of the wide band. 
Note that the $W_1,W_1$ system at $U=4$\,eV is already insulating,
while the $W_2,W_2$ system is much less correlated than the $N_2(\omega)$
spectrum shown in Fig.\,3(a). Conversely, at  $U=6$\,eV the $W_1,W_1$ 
system is far in the insulating range, whereas the  $W_2,W_2$ system is 
close to its metal insulator transition.
  
These predictions can be tested using angle resolved photoemission
spectroscopy. 
In the metallic phase of layered perovskites, for example, the $d_{xy}$ 
and $d_{xz,yz}$ subbands should exhibit coherent peaks of different 
spectral weights as well as different incoherent satellite features. 
At $U_c$ the coherent peaks should vanish simultaneously. In the 
insulating phase, the subbands should exhibit different band gaps 
and Hubbard bands.

For illustrative purposes we have considered the special case of 
half-filled, symmetric subbands. The consistency of the picture discussed 
here with the results obtained for Ca$_{2-x}$Sr$_{x}$RuO$_4$ \cite{EPL} 
suggests that the same conclusion holds for non-symmetric multi-band 
systems. In particular, it should be valid for the Mott insulators 
VO$_2$ and V$_2$O$_3$ and for other transition metal compounds.
   
{\bf Acknowledgements}
 
I am grateful to A. Bringer for discussions and to O. Gunnarsson and
D. Vollhardt for comments. I also like to thank A. Lichtenstein for 
the DMFT-QMC code.

Electronic address: a.liebsch@fz-juelich.de


\begin{thebibliography}{99}
\bibitem{Jbelow1}
Below $U=1$\,eV, $J$ is linearly reduced to zero.
\bibitem{factor}
$U_{c1}$ and $U_{c2}$ do not differ exactly by a factor of two since 
$\beta$ and $J$ are kept constant.
\bibitem{Uc1Uc2}
As shown in Ref.\,16 for the isotropic case, at very low $T$ 
the insulating phase begins at a $U'_c$ slightly below the 
disappearance of the metallic phase. Similar behavior might 
occur for non-isotropic multi-orbital systems.
\end{thebibliography}
\end{document}